\documentclass[prd,aps,showpacs,showkeys,nofootinbib]{revtex4}

\def\XXint#1#2#3{{\setbox0=\hbox{$#1{#2#3}{\int}$}
     \vcenter{\hbox{$#2#3$}}\kern-.5\wd0}}

\usepackage{graphicx}
\usepackage{bm}
\usepackage{amssymb}
\usepackage{amsmath}
\usepackage{euscript}

\begin{document}

\title{Quantum theory of flicker noise: the $1/f$ law as a lower bound on the voltage power spectrum}

\author{Kirill~A.~Kazakov}

\affiliation{Department of Theoretical Physics,
Physics Faculty,\\
Moscow State University, $119991$, Moscow, Russian Federation}

\begin{abstract}
An approach to the problem of $1/f$ voltage noise observed in all conducting media is developed based on an uncertainty relation for the Fourier-transformed signal. It is shown that the quantum indeterminacy caused by non-commutativity of observables at different times sets a lower bound on the power spectrum of voltage fluctuations. Using the Schwinger-Keldysh method, this bound is calculated explicitly in the case of unpolarized free-like charge carriers, and is found to have a $1/f$ low-frequency asymptotic. It is demonstrated also that account of the charge carrier interaction with phonons results in a shift of the frequency exponent from unity. A comparison with the experimental data on $1/f$ noise in InGaAs quantum wells and high-temperature superconductors is made which shows that the observed noise levels are only a few times as high as the bound established.
\end{abstract}

\maketitle

\section{Introduction}\label{introduction}

As is well-known from the experiment, power spectra of the voltage fluctuations in all conducting media exhibit a universal low-frequency behavior: for sufficiently small frequencies $f,$ the spectral density $S(f)\sim 1/f^{\gamma},$ where the frequency exponent $\gamma$ is around unity \cite{johnson, buckingham,bell1980,raychaudhuri}. These ubiquitous fluctuations are often called simply $1/f,$ or flicker, noise.
It has been detected at frequencies as high as $10^6$ Hz down to $10^{-6.3}$ Hz \cite{rollin1953,caloyannides}, with no indication of a low-frequency spectrum flattening. There are various physical processes producing voltage noise: charge carrier trapping-detrapping, motion of dislocations, conductance fluctuations caused by the temperature fluctuations, {\it etc.} The so-called $1/f$ problem can be broadly formulated as a difficulty to relate the $1/f$ spectrum to any of these conventional noise sources. In fact, it is hard to indicate a physical process, say, in a crystal of pure copper, that would be characterized by frequencies much lower than one Hertz. Another side of the problem is the extraordinary wide frequency span of $1/f$-spectrum. In specific experiments, it has often been observed over many decades, whereas the above-mentioned mechanisms produce it only in comparatively narrow frequency bands. For example, defect motion in carbon conductors gives rise to the $1/f$-spectrum only in the range $f \sim 10^2$ Hz--$10^4$ Hz \cite{stephany}. Similarly, if the $1/f$-spectrum observed in a given frequency range is supposed to originate from the charge carrier trapping, the inverse trapping times need to be finely distributed over a much wider range, so that the Lorentzians produced by individual traps would sum up to the observed spectrum. Not saying about quantitative side of the problem (which demands not only the frequency profile, but also the noise magnitude be accounted for), to explain formation of such a perfect distribution is itself a difficult task. In practice, contributions of the conventional noise sources are usually easily identified based on the known material properties and the sample preparation techniques. The {\it fundamental $1/f$ noise} is what remains after these contributions have been eliminated, either analytically in the course of spectrum postprocessing by using appropriate models for the noise mechanisms involved, or experimentally, by improving the sample preparation technology. The $1/f$ problem is thus the question of origin of this minimal-level noise.

The notion of fundamental $1/f$ noise becomes of profound significance once the observed absence of a low-frequency cutoff is accepted as a principle. Until then, subtracting the contributions of known sources from the spectrum is just a convenient way to identify yet another physical process in the conductor. But extending the $1/f$-spectrum down to zero frequency recognizes peculiarity of the fundamental noise, because this leads to an apparent conflict with the observed finiteness of the voltage variance. Indeed, a direct consequence of the familiar Wiener-Khinchin relation \cite{wiener,khinchin} is that this variance is equal to $2\int_{0}^{\infty} {\rm d}f S(f)=\infty.$ It was suggested that the divergence of this integral at $f=0$ is immaterial for $\gamma=1$ (that is, in the case of a logarithmic divergence), in view of the existence of a natural low-frequency cutoff, $f_0,$ -- the inverse lifetime of Universe, which bounds the total power to reasonably moderate values \cite{flinn1968}. Still, this argument does not resolve the problem for $\gamma>1,$ because $\int {\rm d}f S(f) \sim 1/f^{\gamma-1}$ will be unacceptably large for many actual spectra continued down to $f_0.$ In any case, the presumption that the voltage across a small isolated conductor ought to behave differently in a static universe ($f_0=0$) is not quite satisfactory from a purely theoretical standpoint.

Since the Wiener-Khinchin theorem is valid only for processes characterized by stationary autocovariance, one way out of this contradiction is to allow for non-stationarity of the process generating flicker noise. A number of mathematical models has been proposed in this direction which endow electric circuits with certain properties that allow them to act as filters with the response function being a fractional power of frequency, or alternatively, to behave as strange attractors (a survey of this approach can be found in Ref.~\cite{bell1980}, a more recent development, in Refs.~\cite{leibovich2015,dechant2015}). A major issue with these models when applied to the voltage noise is that materials used in microelectronics and related areas do not exhibit desired properties.

An essentially different approach to the problem is taken up in the quantum theory \cite{kazakov1,kazakov2} of fundamental $1/f$ noise according to which this noise is generated by quantum fluctuations of the electromagnetic field produced by free-like charge carriers, rather than by fluctuations in the material properties of the conducting medium. Somewhat more specifically, fluctuations of the electric voltage measured between two probes attached to the sample are correlated via their interaction with the same charge carrier. By this reason, the process can be termed one-particle, though the net effect is a sum of many individual contributions. Thus, the properties of the medium in this picture are important only to the extent that they affect the charge carrier propagation.

In Refs.~\cite{kazakov1,kazakov2}, flicker noise was considered as a finite-temperature effect, namely, the photon heat bath contribution to the power spectrum was only taken into account. The result is proportional to the heat bath temperature, vanishing in the state of photon vacuum. The purpose of the present paper is to demonstrate that the $1/f$ noise actually does not vanish at zero temperature, remaining at a finite level determined by quantum uncertainty in the values of voltage measured at different times. Also, a physical mechanism for the frequency exponent deviation from unity will be described; this deviation was previously taken into account only formally using the techniques of dimensional continuation \cite{kazakov2}. 

The paper is organized as follows. In Sec.~\ref{definition}, the definitions of the voltage autocovariance and power spectrum are discussed from the point of view of Heisenberg's uncertainty. This consideration naturally leads to a general formula for the power spectrum of fundamental noise, given in Sec.~\ref{masterformula}. Technical details for a perturbative calculation of the power spectrum are summarized in Sec.~\ref{perturb}.
The low-frequency asymptotic of the power spectrum is then evaluated in  Sec.~\ref{evaluation} and is shown to be of the type characteristic of the flicker noise. In particular, it is found in Sec.~\ref{lowfrequency} that the strict inverse frequency dependence takes place in the case of freely propagating charge carriers, whereas account of their interaction with phonons, made in Sec.~\ref{gammanonunit}, results in a shift of the frequency exponent from unity. Explicit expressions for the power spectrum are obtained in two measurement setups -- the basic longitudinal and a less common transverse configurations. At last, a comparison with the experimental data in semiconductors and high-temperature superconductors is made in Sec.~\ref{comparison}, and the ensuing conclusions are discussed in Sec.~\ref{discussion} together with the other results obtained.

\section{The noise power spectral density}

\subsection{Autocovariance vs. power spectrum of voltage fluctuations and Heisenberg uncertainty}\label{definition}

Consider a sample with a constant electric current through it, supplied by two leads attached to the sample. For simplicity, the sample material will be assumed macroscopically homogeneous, and so will be the electric field, $\bm{E},$ established inside the sample. Let the voltage across the sample be measured by means of two voltage probes which may or may not coincide with the current leads. Also for simplicity, the probes will be considered pointlike, $\bm{x}_1,\bm{x}_2$ denoting their position. A voltage $U(t,\bm{x}_1,\bm{x}_2)$ measured between the probes at time $t$ is the sum of a constant bias $U_0(\bm{x}_1,\bm{x}_2)$ and a fluctuation, or noise, $\Delta U(t,\bm{x}_1,\bm{x}_2),$
\begin{eqnarray}\label{voltagedecomp}
U(t,\bm{x}_1,\bm{x}_2) = U_0(\bm{x}_1,\bm{x}_2) + \Delta U(t,\bm{x}_1,\bm{x}_2).
\end{eqnarray}
Define a Fourier transform of $\Delta U(t,\bm{x}_1,\bm{x}_2)$ measured during time $t_m$ (omitting, for brevity, the arguments $\bm{x}_1,\bm{x}_2$)
\begin{eqnarray}\label{fourier}
\Delta U_s(\omega) = \int_{0}^{t_m}{\rm d}t\, \Delta U(t)\sin(\omega t),
\quad \Delta U_c(\omega) = \int_{0}^{t_m}{\rm d}t\, \Delta U(t)\cos(\omega t).
\end{eqnarray} In contrast to Refs.~\cite{kazakov1,kazakov2}, it is written in a real form which is more suitable for the subsequent discussion. In actual experimental signal processing, a discrete Fourier transform is used to evaluate $\Delta U_{s,c}(\omega)$, according to the discreteness of the signal recording. After this measurement has been repeated many times, the main quantity of interest -- the noise power spectral density -- is obtained as
\begin{eqnarray}\label{power}
S(f) = \lim\limits_{t_m\to \infty}\frac{1}{t_m}\left\langle \left(\Delta U_s(\omega)\right)^2 + \left(\Delta U_c(\omega)\right)^2\right\rangle, \quad \omega = 2\pi f,
\end{eqnarray} where the angular brackets denote averaging over the set of signal samplings, and $t_m\to\infty$ designates the condition that the measurement time be much larger than the inverse of given frequency.

As discussed in Sec.~\ref{introduction}, a stumbling block of the $1/f$ problem is the divergence of the total noise power at zero frequency, which is apparently suggested by the Wiener-Khinchin theorem. To briefly recall the argument, assume that the random process $\Delta U(t)$ is wide-sense stationary, that is, its autocovariance $\langle \Delta U(t)\Delta U(t+\tau)\rangle \equiv C(\tau)$ is a function of the lag $\tau$ only. This function is then even, for $$C(-\tau)=\langle \Delta U(t)\Delta U(t-\tau)\rangle = \langle \Delta U(t+\tau)\Delta U(t)\rangle = \langle \Delta U(t)\Delta U(t+\tau)\rangle = C(\tau),$$ where the third equality follows from commutativity of $c$-functions $\Delta U(t)$ and $\Delta U(t+\tau).$ The Wiener-Khinchin theorem states that for the considered process,
\begin{eqnarray}\label{wiener}
C(\tau) = \int_{-\infty}^{+\infty}{\rm d}fS(f){\rm e}^{{\rm i}2\pi f\tau}\,,
\end{eqnarray} provided that $C(\tau)$ is finite for all $\tau.$ Since the evenness of $C(\tau)$ implies that of $S(f),$ the variance of $\Delta U$ is related to the total power as $$\langle (\Delta U(t))^2\rangle = C(0)=2\int_{0}^{+\infty}{\rm d}fS(f),$$ which diverges for $S(f)\sim 1/f.$ 

In this consideration, $\Delta U(t)$ is a random variable which is classical in that it takes on a real value at any given $t$ (a voltmeter reading). In the case of a quantum underlying process, in order to evaluate the function $C(\tau)$ one needs to construct a quantum-mechanical operator that would represent the product $\Delta U(t)\Delta U(t+\tau).$ From the quantum theory standpoint, the voltage $U(t)$ is an observable to which there corresponds a Hermitian (Heisenberg) operator $\widehat{U}(t).$ For a given bias $U_0(\bm{x}_1,\bm{x}_2)\equiv U_0$, this defines another observable -- the voltage fluctuation -- by virtue of Eq.~(\ref{voltagedecomp}),
\begin{eqnarray}\label{voltagedecomp1}
\widehat{\Delta U}(t) = \widehat{U}(t) - U_0.
\end{eqnarray}
But one cannot construct an observable for the product $\Delta U(t)\Delta U(t+\tau)$ by multiplying $\widehat{\Delta U}(t)$ and $\widehat{\Delta U}(t+\tau),$ because these operators do not commute with each other, so that their multiplication is ambiguous. As is well known, the physical meaning of this non-commutativity is that $\Delta U$ cannot have definite values at two instants $t,t+\tau$ in any state with a nonzero expectation value of the commutator $$\widehat{\Delta U}(t)\widehat{\Delta U}(t+\tau) - \widehat{\Delta U}(t+\tau)\widehat{\Delta U}(t) \equiv [\widehat{\Delta U}(t),\widehat{\Delta U}(t+\tau)].$$ As a result, $\Delta U$ is smeared at every instant over a set of values, each of which can be detected with some probability upon measurement. Put somewhat differently, this means that it is impossible to obtain definite values of $\Delta U$ at two different instants without changing the system state: even if the state vector of the system ``conducting sample plus electromagnetic field'' was changed negligibly during the first measurement with a definite outcome, it will necessarily undergo a finite change during a later measurement that results in a definite $\Delta U.$ Since the set of all possible values of $\Delta U$ is a continuum, the term ``definite'' means, as usual, that the measurement yields $\Delta U$ in a sufficiently narrow interval determined by the measurement precision. It follows that each sampling of $\Delta U$ during the time $t_m$ is accompanied by a continuous alteration of the system state in a way consistent with the uncertainty principle: a higher measurement precision entails larger system state variations, hence larger voltage fluctuations between successive measurements. The voltage autocovariance thus depends on the way the system state varies during the measurement. As a result of the temporal inhomogeneity caused by this intervention of the measuring device, the autocovariance function will generally depend on both arguments $t$ and $(t+\tau),$ not only on their difference. An alternative we thus face is that {\it either, for a fixed system state, the voltage fluctuation is not defined for all times in principle, or $\left\langle\widehat{\Delta U}(t)\widehat{\Delta U}(t+\tau)\right\rangle$ is $t$-dependent.} But since the Wiener-Khinchin theorem assumes that the process $\Delta U(t),$ though random, is both well-defined for all $t$ and is stationary, it does not apply in either case.

Though it is not the subject of the present paper, it is worth mentioning that in principle, the dependence of autocovariance on the measuring device operation can be explicitly described using the method of projection operators, or alternatively, a restricted path integral (see, {\it e.g.} Refs.~\cite{ballentine,mensky}). It is to be emphasized that this conditioning on the measurement procedure not only makes the autocovariance dependent on both time arguments, but also introduces a true asymmetry between the instants $t$ and $(t+\tau),$ because the measurement of $\Delta U(t)$ affects the system state, hence the results of the measurement at a later instant $(t+\tau),$ but not vice versa.

It is important, on the other hand, that the quantum-mechanical description of the power spectrum is free of these complications. To be sure, the non-commutativity of $\widehat{\Delta U}(t)$ and $\widehat{\Delta U}(t+\tau)$ is still there, but it is now inconsequential. To see this, we first note that $\widehat{\Delta U}(t)$ uniquely defines two other Hermitian operators $\widehat{\Delta U_s}(\omega),$ $\widehat{\Delta U_c}(\omega)$ according to Eq.~(\ref{fourier})
\begin{eqnarray}\label{fourier1}
\widehat{\Delta U_s}(\omega) = \int_{0}^{t_m}{\rm d}t\, \widehat{\Delta U}(t)\sin(\omega t),
\quad \widehat{\Delta U_c}(\omega) = \int_{0}^{t_m}{\rm d}t\, \widehat{\Delta U}(t)\cos(\omega t),
\end{eqnarray} and that the squares of these are also uniquely defined despite the non-commutativity of $\widehat{\Delta U}(t)$ at different $t$s. Indeed, one has
\begin{eqnarray}
\left(\widehat{\Delta U_s}(\omega)\right)^2 &=& \iint_{0}^{t_m}{\rm d}t{\rm d}t'\, \widehat{\Delta U}(t) \widehat{\Delta U}(t')\sin(\omega t)\sin(\omega t') \nonumber\\ &=& \iint_{0}^{t_m}{\rm d}t{\rm d}t'\, \frac{1}{2}\left(\widehat{\Delta U}(t)\widehat{\Delta U}(t')+\widehat{\Delta U}(t')\widehat{\Delta U}(t)\right)\sin(\omega t)\sin(\omega t'),
\end{eqnarray} and similarly for $(\widehat{\Delta U_c}(\omega))^2.$ That is, these operators involve only the symmetrized product of $\widehat{\Delta U}(t)$ at different times.
And second, the definition of $S(f)$ according to Eq.~(\ref{power})  exactly corresponds to the usual quantum-mechanical formula for calculating averages, namely,
\begin{eqnarray}\label{powerq}
S(f) = \lim\limits_{t_m\to \infty}\frac{1}{t_m}\left\langle \Psi \left| \left(\widehat{\Delta U_s}(\omega)\right)^2 + \left(\widehat{\Delta U_c}(\omega)\right)^2\right|\Psi\right\rangle,
\end{eqnarray} where $|\Psi\rangle $ is the state vector of the system. In fact, it is a definition of the observable $\hat{f}$ that $\langle\Psi| \hat{f}|\Psi\rangle$ equals the average of a set of values obtained in a series of repeated measurements of the physical quantity $f$ in the given state $\Psi$. Here, the non-commutativity of $\widehat{\Delta U}(t)$ at different $t$s shows itself as a non-commutativity of the two operators $\left(\widehat{\Delta U_s}(\omega)\right)^2$ and $\left(\widehat{\Delta U_c}(\omega)\right)^2$, which, however, causes presently no ambiguity, because the average of the sum is equal to the sum of averages.

Thus, we arrive at the conclusion that in contrast to the autocovariance, the power spectrum of voltage fluctuations can be self-consistently computed using the formula (\ref{powerq}). Since in practice one normally deals with systems close to thermodynamic equilibrium, the vector $|\Psi\rangle $ appearing in this formula can be taken as one of the vectors representing given macroscopic state of the system at a given temperature, and then the usual statistical averaging over an ensemble of such states brings Eq.~(\ref{powerq}) to
\begin{eqnarray}\label{powerq1}
S(f) = \lim\limits_{t_m\to \infty}\frac{1}{t_m}\left\{\left\langle  \left(\widehat{\Delta U_s}(\omega)\right)^2\right\rangle + \left\langle\left(\widehat{\Delta U_c}(\omega)\right)^2\right\rangle\right\},
\end{eqnarray} where $\left\langle \hat{f} \right\rangle = {\rm tr}\left( \hat{\rho}\hat{f}\right),$ $\hat{\rho}$ being the system density matrix.

\subsection{General formula for the power spectrum of fundamental  noise}\label{masterformula}

As described in Sec.~\ref{introduction}, the fundamental $1/f$ noise is due to residual fluctuations that remain in the system upon elimination of the conventional sources that generate noise by affecting the sample conductivity. Following this characterization, it is natural to seek for the minimal level of voltage fluctuations. According to the discussion in Sec.~\ref{definition}, this level cannot be lower than that allowed by the uncertainty principle. Therefore, in order to determine the lower bound on the noise level, expression (\ref{powerq1}) is to be minimized with the understanding that quantum indeterminacy is the only source of voltage fluctuations. Such bound, if any, is naturally expected to be independent of the measurement process specifics. Therefore, we can exclude the measuring device from consideration by assuming that $\hat{\rho}$ describes a {\it fixed} state of the system ``conducting sample plus electromagnetic field.'' An uncertainty relation for the two observables $\Delta U_s(\omega), \Delta U_c(\omega)$ then reads
\begin{eqnarray}\label{indeterm}
\left\langle \left(\widehat{\Delta U_s}(\omega)\right)^2\right\rangle \left\langle \left(\widehat{\Delta U_c}(\omega)\right)^2\right\rangle \geqslant \frac{1}{4}\left| \left\langle \left[\widehat{\Delta U_s}(\omega),\widehat{\Delta U_c}(\omega)\right]\right\rangle\right|^2.
\end{eqnarray} Introducing notation
\begin{eqnarray}\label{powerf}
S_F(f)&=&\lim\limits_{t_m\to \infty}\frac{1}{t_m}\left| \left\langle \left[\widehat{\Delta U_s}(\omega),\widehat{\Delta U_c}(\omega)\right]\right\rangle\right|
\end{eqnarray} and
\begin{eqnarray}
D_{s,c}(f)&=&\lim\limits_{t_m\to \infty}\frac{1}{t_m}\left\langle \left(\widehat{\Delta U_{s,c}}(\omega)\right)^2 \right\rangle,\nonumber
\end{eqnarray}
we combine Eqs.~(\ref{powerq}), (\ref{indeterm}) into
$$S\geqslant D_s+\frac{S_F^2}{4D_s}\,.$$ The right hand side of this inequality takes on a minimal value $S_F$ at $D_s=S_F/2,$ so that also $D_c=S_F/2.$ Therefore, {\it we identify expression (\ref{powerf}) as the power spectrum of fundamental noise.}

An important observation to be made is that the expectation value of the commutator on the right hand side of Eq.~(\ref{powerf}) is an {\it odd} function of frequency, for $\widehat{\Delta U_s}(\omega)$ is odd, whereas $\widehat{\Delta U_c}(\omega)$ is even in $\omega$. More specifically, using Eq.~(\ref{fourier1}) and interchanging the integration variables, one has
\begin{eqnarray}\label{commutator}
\left\langle\left[\widehat{\Delta U_s}(\omega),\widehat{\Delta U_c}(\omega)\right]\right\rangle &=& \iint_{0}^{t_m}{\rm d}t{\rm d}t'\, \left\langle[\widehat{\Delta U}(t),\widehat{\Delta U}(t')]\right\rangle\sin(\omega t)\cos(\omega t') \nonumber\\ &=& \iint_{0}^{t_m}{\rm d}t{\rm d}t'\, \left\langle\widehat{\Delta U}(t)\widehat{\Delta U}(t')\right\rangle\sin(\omega (t - t')).
\end{eqnarray} It follows that the power spectrum is determined by the Fourier transform of the contribution to $\left\langle\widehat{\Delta U}(t)\widehat{\Delta U}(t')\right\rangle$ which is antisymmetric under the interchange $t\leftrightarrow t'.$ Thus, when computing this expectation value, one can safely use the standard momentum-space techniques of quantum field theory to extract its low-frequency asymptotic despite anticipated zero-frequency singularity, for unlike an even-frequency function $1/|\omega|,$ Fourier transform of $1/\omega$ does exist. On the other hand, $S(f)$ is determined by the symmetric part of $S(t-t'),$
$$\left\langle  \left(\widehat{\Delta U_s}(\omega)\right)^2\right\rangle + \left\langle\left(\widehat{\Delta U_c}(\omega)\right)^2\right\rangle = \iint_{0}^{t_m}{\rm d}t{\rm d}t'\, \left\langle\widehat{\Delta U}(t)\widehat{\Delta U}(t')\right\rangle\cos(\omega (t - t')),$$
and this part lacks Fourier decomposition when $S(f)\sim 1/|\omega|.$
We see that while $S_F(f)$ is derived from a function admitting Fourier decomposition, Fourier transform of $S_F(f)$ itself may not exist. In other words, the right hand side of Eq.~(\ref{wiener}) does not have to be well-defined, and as we have seen, neither does its left hand side.

In an open system such as the sample conducting electric current, $\left\langle\widehat{\Delta U}(t)\widehat{\Delta U}(t')\right\rangle$ does not have to depend on the difference $(t-t')$ only, even though the influence of the measuring device is negligible and the external conditions are stationary. But because of the factor $1/t_m$ in Eq.~(\ref{powerf}), it is terms dependent solely on $(t-t')$ that only contribute in the limit $t_m\to \infty.$ Let us distinguish them with a minus subscript, $\left\langle\widehat{\Delta U}(t)\widehat{\Delta U}(t')\right\rangle_- \equiv S(t-t'),$ and consider the quantity
$$\Sigma(f) = \lim\limits_{t_m\to \infty}\frac{1}{t_m}\iint_{0}^{t_m}{\rm d}t{\rm d}t'\, S(t-t') {\rm e}^{{\rm i}\omega (t - t')}.$$ On applying Euler's formula to the exponent in the integrand, the contribution of cosine to $\Sigma(f)$ is just $S(f),$ whereas the modulus of the other contribution is $S_F(f).$ Writing 
$$\iint_{0}^{t_m}{\rm d}t{\rm d}t'\, S(t-t') {\rm e}^{{\rm i}\omega (t - t')} = \int_{0}^{t_m}{\rm d}t'\int_{-t'}^{t_m-t'}{\rm d}\tau S(\tau) {\rm e}^{{\rm i}\omega\tau}$$ and integrating by parts, this can be reduced to
\begin{eqnarray}\label{sigma}
\Sigma(f) = \lim\limits_{t_m\to \infty}\left\{\int_{-t_m}^{t_m}{\rm d}\tau S(\tau) {\rm e}^{{\rm i}\omega\tau} - \frac{1}{t_m}\int_{-t_m}^{t_m}{\rm d}\tau |\tau|S(\tau) {\rm e}^{{\rm i}\omega\tau}\right\}.
\end{eqnarray} If one assumes that $S(\tau)$ sufficiently rapidly decreases as $|\tau|\to \infty,$ the second term vanishes in the limit $t_m\to\infty,$ and Eq.~(\ref{sigma}) takes the same form as the conventional Wiener-Khinchin relation for processes with a well-defined autocovariance. But the $1/f$ noise is just the case where this assumption does not hold. The $1/|\omega|$ asymptotic of the power spectrum suggests that at large $|\tau|$'s, $S(\tau)$ contains a term $\sim \ln|\tau|.$ It is not difficult to check that
\begin{eqnarray}\label{integrals}
\int_{-t_m}^{t_m}{\rm d}\tau \ln|\tau| {\rm e}^{{\rm i}\omega\tau} &=& -\frac{\pi}{|\omega|} - 2\ln t_m\frac{\sin(\omega t_m)}{\omega}  + O(1/t_m), \nonumber \\
\frac{1}{t_m}\int_{-t_m}^{t_m}{\rm d}\tau |\tau|\ln|\tau| {\rm e}^{{\rm i}\omega\tau} &=& - 2\ln t_m\frac{\sin(\omega t_m)}{\omega}  + O(1/t_m).\end{eqnarray} It is seen that despite these expressions diverge in the limit $t_m\to\infty,$ their difference converges to $-\pi/|\omega|.$ It follows that the right hand side of Eq.~(\ref{sigma}) is also well-defined for the function $S(\tau) = \ln[a + (\tau/\tau_0)^2]$ with constant $\tau_0,a>0,$ and that $\Sigma(f) \to -1/|f|$ for $|f|\ll 1/\tau_0.$ This simple example well illustrates the above statement that the power spectrum may exist even for processes for which $S(t-t')$ does not admit Fourier decomposition, and that the presence of a $1/f$-term in its low-frequency asymptotic does not contradict finiteness of the voltage variance. In fact, in the present example, $S(0) = \ln a$ can be any real number despite the ``total power'' $\int_{-\infty}^{\infty}{\rm d}f\Sigma(f)$ is infinite.

\section{Perturbation theory for calculating the power spectrum}\label{perturb}

We proceed to the calculation of the right hand side of Eq.~(\ref{powerf}) using perturbation theory. For this purpose, the Schwinger-Keldysh technique \cite{schwinger,keldysh} for computing in-in expectation values will be used. In contrast to Refs.~\cite{kazakov1,kazakov2}, no attempt will be made to simplify this technique by relating the in-in expectation values to the corresponding in-out matrix elements, however natural such relation might appear.
To make the presentation self-contained, the notation and main formulas of the Schwinger-Keldysh technique will be briefly recalled first. To simplify intermediate expressions, we switch henceforth to units in which $\hbar = c =1,$ $c$ denoting the speed of light in vacuum.

\subsection{The Schwinger-Keldysh rules}

In order to evaluate the power spectrum, the expectation value
\begin{eqnarray}\label{expectation}
\left\langle\widehat{\Delta U}(t)\widehat{\Delta U}(t')\right\rangle \equiv {\rm tr}\left(\hat{\rho}\widehat{\Delta U}(t)\widehat{\Delta U}(t')\right)
\end{eqnarray}
appearing on the right of Eq.~(\ref{commutator}) is to be brought to a form suitable for perturbative expansion. A standard procedure is essentially as follows. The total Hamiltonian $\hat{H}$ is first split into a free part $\hat{H}_0$ and an interaction $\hat{W},$ $\hat{H}= \hat{H}_0 + \hat{W}.$ As discussed in Sec.~\ref{definition}, the power spectrum of fundamental noise can be computed with no reference to the measurement process, and so $\hat{W}$ does not include the energy of interaction with the measuring device. $\hat{H}$ and $\hat{H}_0$ determine the time dependence of Heisenberg and interaction-picture operators, respectively (the latter are denoted by lower-case letters), {\it viz.},
$$\widehat{\Delta U}(t) = \exp({\rm i}t\hat{H})\widehat{\Delta U}(0)\exp(-{\rm i}t\hat{H}), \quad \widehat{\Delta u}(t) = \exp({\rm i}t\hat{H}_0)\widehat{\Delta U}(0)\exp(-{\rm i}t\hat{H}_0),$$
$$\hat{w}(t) = \exp({\rm i}t\hat{H}_0)\hat{W}\exp(-{\rm i}t\hat{H}_0),$$ {\it etc.} Using the formula
$$\exp({\rm i}t\hat{H}_0)\exp(-{\rm i}t\hat{H}) = \hat{S}(t,t_0)\exp({\rm i}t_0\hat{H}_0)\exp(-{\rm i}t_0\hat{H}), \quad t>t_0,$$ where
$$\hat{S}(t,t_0) = \EuScript{T}\exp\left\{-{\rm i}\int_{t_0}^{t}{\rm d}t\,\hat{w}(t)\right\},$$ and $\EuScript{T}$ denotes a chronological ordering of the operators $\hat{w}(t)$ such that their arguments $t$ increase leftward, the expectation value (\ref{expectation}) can be written as
\begin{eqnarray}\label{expectation1}
{\rm tr}\left(\hat{\rho}_0\hat{S}^{-1}(t,t_0)\widehat{\Delta U}_0(t)\hat{S}(t,t_0)\hat{S}^{-1}(t',t_0)\widehat{\Delta U}_0(t')\hat{S}(t',t_0)\right),
\end{eqnarray} where
$$\hat{\rho}_0 \equiv \exp({\rm i}t_0\hat{H}_0)\exp(-{\rm i}t_0\hat{H})\hat{\rho}\exp({\rm i}t_0\hat{H})\exp(-{\rm i}t_0\hat{H}_0).$$
As usual, an adiabatic switching on of the quasi-particle interactions from $t = -\infty$ to finite times can be used to specify any given system state $\left|\Psi\right\rangle$ as a superposition of the so-called in-states, that is states wherein the charge carriers and photons (as well as other quasi-particles such as phonons, if any) are freely propagating in the remote past, all with definite momenta and spin/polarizations. The operator $\exp({\rm i}t_0\hat{H}_0)\exp(-{\rm i}t_0\hat{H})$ then has the virtue of transforming, in the limit $t_0 \to -\infty,$ any in-state $\left|\Psi,{\rm in}\right\rangle$ into the corresponding state $\left|\Psi_0\right\rangle$ of non-interacting particles with the given momenta and spin/polarizations, according to $$\left|\Psi_0\right\rangle = \lim\limits_{t_0 \to -\infty}\exp({\rm i}t_0\hat{H}_0)\exp(-{\rm i}t_0\hat{H})\left|\Psi,{\rm in}\right\rangle.$$ In fact, this relation just asserts that the states $\left|\Psi,{\rm in}\right\rangle$ and $\left|\Psi_0\right\rangle$ get the same appearance once translated into remote past by the operators $\exp(-{\rm i}t_0\hat{H})$ and $\exp(-{\rm i}t_0\hat{H}_0),$ respectively. Thus, $\hat{\rho}_0$ is the density matrix of the system wherein the charge carriers and photons are in a mixture of free particle states with the given distribution over momenta and spin/polarizations.

Next, setting $t_0 = -\infty$ in expression (\ref{expectation1}), using the relation $$\hat{S}(t,-\infty)\hat{S}^{-1}(t',-\infty) = \hat{S}^{-1}(+\infty,t)\hat{S}(+\infty,t'),$$ and noting that the operators $\hat{w}(t)$ in $\hat{S}^{-1}$ are ordered anti-chronologically (that is, the time arguments in a product of $\hat{w}(t)$s increase rightward), the expectation value (\ref{expectation}) can be written as
\begin{eqnarray}\label{expectation2}
\left\langle\widehat{\Delta U}(t)\widehat{\Delta U}(t')\right\rangle = {\rm tr}\left(\hat{\rho}_0\EuScript{T}_C\widehat{\Delta u}^{(2)}(t)\widehat{\Delta u}^{(1)}(t')\exp\left\{-{\rm i}\int_{C}{\rm d}t\,\hat{w}(t)\right\}\right),
\end{eqnarray}
where the so-called Schwinger-Keldysh time contour $C$ runs from $t=-\infty$ to $t=+\infty$, and then back to $t=-\infty,$ with the forward branch designated with a superscript $(1)$ and treated as being in the past with respect to any instant on the backward branch designated with a superscript $(2).$ $\EuScript{T}_C$ accordingly orders all operators along the contour $C$: it is  chronological (the usual $T$-ordering) on the forward branch of $C$, and anti-chronological on its backward branch. In Eq.~(\ref{expectation2}),  $\widehat{\Delta u}(t)$ and $\widehat{\Delta u}(t')$ are assigned the superscripts $(2)$ and $(1),$ respectively, because their product on the left is not time-ordered, $\widehat{\Delta U}(t)$ standing to the left of $\widehat{\Delta U}(t')$ for all $t,t'.$

The right hand side of Eq.~(\ref{expectation2}) is now suitable for a perturbative expansion with respect to whatever couplings appear in $\hat{w}(t).$ We shall first consider the case where a single-species charge carriers and photons are the only free-like particles present in the system. Interaction of the charge carriers with phonons will be taken into account later on in Sec.~\ref{gammanonunit}. Furthermore, all thermal effects will be neglected, in particular, those related to the photon heat bath. Then the system density matrix $\hat{\rho}_0$ reduces to that of the charge carriers, $\hat{\varrho},$ times the photon vacuum. Although a fully relativistic treatment can be developed, for most practical applications it is sufficient to consider the charge carriers as non-relativistic. In fact, as the results of Refs.~\cite{kazakov1,kazakov2} show, the fundamental $1/f$ noise is a relativistic phenomenon only to the extent that such is the photon propagation. By this reason, the charge carriers will be assumed non-relativistic fermions with the electric charge $e$ and effective mass $m.$ Also for simplicity, they will be taken unpolarized, and all spin indices will be suppressed throughout. Therefore, an interaction-picture fermion field operator is
\begin{eqnarray}\label{fermiondecomp}
\hat{\phi}(t,\bm{x}) = \int_{-\infty}^{+\infty}\frac{{\rm d}^3\bm{q}}{(2\pi)^3}\hat{b}_{\bm{q}}\exp(-{\rm i}\varepsilon_{\bm{q}}t + {\rm i}\bm{q}\cdot\bm{x}),
\end{eqnarray}
where $\varepsilon_{\bm{q}}$ is the energy of charge carrier with momentum $\bm{q},$ and the destruction operators $\hat{b}_{\bm{q}}$ satisfy $\hat{b}_{\bm{q}}\hat{b}^{\dagger}_{\bm{q}'} + \hat{b}^{\dagger}_{\bm{q}'}\hat{b}_{\bm{q}} = (2\pi)^3\delta^{(3)}({\bm{q} - \bm{q}'}).$
As usual, on expanding the right hand side of Eq.~(\ref{expectation2}) in powers of the electromagnetic coupling and using the macroscopic Wick theorem, it is expressed as the sum of products of particle propagators. A complication that makes the calculation of expectation values so laborious compared to the Feynman techniques for calculating in-out matrix elements is that there are four propagators for each particle, depending on which branch of the contour $C$ the time arguments of the propagator belong to. Thus, the fermion propagator is a matrix
\begin{eqnarray}\label{fermionprop}
D^{(ij)}(x,y) = {\rm tr}\left(\hat{\varrho}\EuScript{T}_C \hat{\phi}^{(i)}(x^0,\bm{x})\hat{\phi}^{(j)\dagger}(y^0,\bm{y})\right),
\end{eqnarray}
where indices $i,j$ take on values $1,2$, and the action of $\EuScript{T}_C$ is extended over single-field operators by specifying that an interchange of two fermionic operators is accompanied by a factor of $(-1).$ Likewise, the photon propagator matrix is
\begin{eqnarray}\label{photonprop}
G^{(ij)}_{\mu\nu}(x,y) = \left\langle 0 | \EuScript{T}_C \hat{a}_{\mu}^{(i)}(x^0,\bm{x})\hat{a}_{\nu}^{(j)}(y^0,\bm{y})|0\right\rangle, \quad \mu,\nu=0,1,2,3,
\end{eqnarray}
where $\hat{a}_{\mu}$ is the interaction-picture operator for the electromagnetic 4-potential $A_{\mu}$, and the vacuum mean is taken according to the above choice of the photon state. The products of particle propagators are integrated over time with appropriate vertex factors generated by the interaction Hamiltonian $\hat{w}(t)$, each vertex belonging to either branch of the contour $C.$ It is convenient to let all time integration variables run from $-\infty$ to $+\infty$; then the vertices on the backward branch will carry an extra factor of $(-1).$ Otherwise, the rules of perturbative calculation of the right hand side of (\ref{expectation2}) are the same as the Feynman rules for calculating the two-point Green functions in the scattering theory.

\subsection{Particle propagators in momentum space}\label{oneparticle}

Momentum-space expressions for the propagators are obtained by substituting the normal-mode decompositions of the field operators. The electromagnetic field decomposition reads
$$\hat{a}_{\mu}(x) = \int_{-\infty}^{+\infty}\frac{{\rm d}^3 \bm{k}}{(2\pi)^3}\sum\limits_{a}\frac{\sqrt{4\pi}}{\sqrt{2|\bm{k}|}}\left(\hat{c}_{\bm{k}a}e_{\mu a}(\bm{k}){\rm e}^{-{\rm i}kx} + \hat{c}^{\dagger}_{\bm{k}a}e^*_{\mu a}(\bm{k}){\rm e}^{{\rm i}kx}\right)_{k^0 = |\bm{k}|},$$ with the operators $\hat{c}_{\bm{k}a}$ satisfying $\hat{c}_{\bm{k}a}\hat{c}^{\dagger}_{\bm{k}'b} - \hat{c}^{\dagger}_{\bm{k}'b}\hat{c}_{\bm{k}a} = \delta_{ab}(2\pi)^3\delta^{(3)}({\bm{k} - \bm{k}'}),$ and Latin subscripts $a,b$ numbering photon polarizations $e_{\mu}(\bm{k}).$ Substitution into Eq.~(\ref{photonprop}) yields
\begin{eqnarray}\label{photonprop1}
G^{(ij)}_{\mu\nu}(x,y) &=& \int_{-\infty}^{+\infty}\frac{{\rm d}^3\bm{k}}{(2\pi)^3}\frac{{\rm d}k^0}{2\pi}P_{\mu\nu}(k)G^{(ij)}(k){\rm e}^{-{\rm i}k^0(x^0-y^0) + {\rm i}\bm{k}\cdot(\bm{x}-\bm{y})} \nonumber\\ &\equiv& \int_{-\infty}^{+\infty}\frac{{\rm d}^4 k}{(2\pi)^4}P_{\mu\nu}(k)G^{(ij)}(k){\rm e}^{-{\rm i}k(x-y)}, \quad P_{\mu\nu}(k) = 4\pi\sum\limits_{a} e^*_{\mu a}(\bm{k})e_{\nu a}(\bm{k}),\nonumber\\
G^{(11)}(k) &=& \frac{{\rm i}}{k^2 + {\rm i}0} = [G^{(22)}(k)]^*, \quad G^{(12)}(k) = 2\pi\theta(-k^0)\delta(k^2), \quad G^{(21)}(k) = 2\pi\theta(k^0)\delta(k^2),\nonumber
\end{eqnarray} where $\theta(k^0)$ is the step function [$\theta(k^0)=0$ for $k^0 \leqslant 0$, $\theta(k^0)=0$ for $k^0>0$]. Hereon, the space-time or momentum-energy coordinates and integrals are written as ``4-dimensional,'' $x=(x^0,\bm{x}), k=(k^0,\bm{k}),$ $kx = k^0x^0 - \bm{k}\cdot\bm{x}.$ 

Likewise, substitution of the fermion field decomposition (\ref{fermiondecomp}) into Eq.~(\ref{fermionprop}) gives the charge carrier propagator as the sum
$$D^{(ij)}(x,y) = D^{(ij)}_{0}(x,y) + \Delta(x,y)$$
of a vacuum part
\begin{eqnarray}\label{fermionprop1}
D^{(ij)}_{0}(x,y) &=& \int_{-\infty}^{+\infty}\frac{{\rm d}^4 q}{(2\pi)^4}D^{(ij)}_{0}(q){\rm e}^{-{\rm i}q(x-y)}, \nonumber\\
D^{(11)}_{0}(q) &=& \frac{{\rm i}}{q^0 - \varepsilon_{\bm{q}} + {\rm i}0} = [D^{(22)}_{0}(q)]^*, \quad \nonumber\\ D^{(12)}_{0}(q) &=& 0, \quad D^{(21)}_{0}(q) = 2\pi\delta(q^0 - \varepsilon_{\bm{q}}),
\end{eqnarray}
and a part dependent on the charge carrier distribution,
\begin{eqnarray}\label{delta}
\Delta(x,y) &=& \int_{-\infty}^{+\infty}\frac{{\rm d}^3 \bm{q}}{(2\pi)^3}\frac{{\rm d}^3 \bm{q}'}{(2\pi)^3}\varrho(\bm{q},\bm{q}'){\rm e}^{-{\rm i}(\varepsilon_{\bm{q}}x^0 - \varepsilon_{\bm{q}'}y^0) +
{\rm i}(\bm{q}\bm{x} - \bm{q}'\bm{y})},
\end{eqnarray}
where
\begin{eqnarray}\label{chdensitym}
\varrho(\bm{q},\bm{q}') = {\rm tr}\left(\hat{\varrho}\hat{b}^{\dagger}_{\bm{q}'}\hat{b}_{\bm{q}}\right)
\end{eqnarray} is the momentum-space density matrix of the charge carriers. It is to be noted that this matrix is not diagonal, because charge carriers are bound to pass through the sample of finite dimensions. Its normalization will be discussed later in Sec.~\ref{normalization}.

As the elements of Feynman diagrams, $D_0$ and $\Delta$ will be depicted by solid and broken lines, respectively, whereas wavy lines will represent photon propagators. In momentum space, $D_0$ and $G_{\mu\nu}$ are functions of the corresponding particle 4-momenta, in contrast to $\Delta$ which depends on a pair of 3-momenta $\bm{q},\bm{q}'$, or in 4-dimensional notation, on the charge carrier 4-momenta on the mass shell $q = (\varepsilon_{\bm{q}},\bm{q})$ and $q' = (\varepsilon_{\bm{q}'},\bm{q}')$.

\subsection{Feynman gauge. Formula (\ref{expectation2}) for non-relativistic charge carriers.}

Representing an observable quantity, the noise spectral density (\ref{powerf}) is independent of the gauge condition which is necessarily imposed on the fields in intermediate calculations to fix the gradient invariance of electrodynamics. At the same time, the calculational complexity directly depends on the gauge, and a properly chosen gauge can greatly simplify computations. In the present context, the gauge choice is suggested by the non-relativistic nature of charge carriers.

From the experimental standpoint, the voltage measurement amounts to measuring small currents through a high-resistance wire connecting the voltage probes, hence to measuring the value of the integral
\begin{eqnarray}\label{integral}
\int_{\bm{x}_1}^{\bm{x}_2} {\rm d}\bm{r}\cdot\bm{E}(t,\bm{r}) = U(t,\bm{x}_1,\bm{x}_2)
\end{eqnarray}
taken along the wire, where $\bm{E}(t,\bm{r}) = -\bm{\nabla}A_0 - \partial\bm{A}/\partial t$ is the fluctuating electric field. The corresponding quantum-mechanical expression for $\langle \Delta U(t)\Delta U(t')\rangle$ is obtained by substituting the voltage fluctuation [Cf. Eq.~(\ref{voltagedecomp})], \begin{eqnarray}\label{fluctuation}
\Delta U(t,\bm{x}_1,\bm{x}_2) = \int_{\bm{x}_1}^{\bm{x}_2} {\rm d}\bm{r}\cdot\bm{E}(t,\bm{r}) - U_0
\end{eqnarray} into Eq.~(\ref{expectation2}):
\begin{eqnarray}\label{expectation3}
&&\left\langle\widehat{\Delta U}(t)\widehat{\Delta U}(t')\right\rangle \nonumber \\&& = {\rm tr}\left(\hat{\rho}_0\EuScript{T}_C\int_{\bm{x}_1}^{\bm{x}_2} {\rm d}\bm{r}\cdot\hat{\bm{e}}^{(2)}(t,\bm{r})
\int_{\bm{x}_1}^{\bm{x}_2} {\rm d}\bm{r}'\cdot\hat{\bm{e}}^{(1)}(t',\bm{r}')\exp\left\{-{\rm i}\int_{C}{\rm d}t\,\hat{w}(t)\right\}\right) - U^2_0,
\end{eqnarray} where $\hat{\bm{e}} = -\bm{\nabla}\hat{a}_0 - \partial\hat{\bm{a}}/\partial t$ is the electric field operator in the interaction picture.

In the absence of strong time-varying magnetic fields, which is normally the case in flicker noise studies, integral (\ref{integral}) is independent of the wire shape (provided that its ends are fixed on the voltage probes), but its expression in terms of the 4-potential $A^{\mu}$ does depend on the gauge. Also gauge-dependent is the polarization matrix $P_{\mu\nu}$ appearing in the photon propagator (\ref{photonprop1}). As is well known, this matrix can be taken proportional to the Minkowski tensor $\eta_{\mu\nu} = {\rm diag (1,-1,-1,-1)},$
\begin{eqnarray}\label{feynmangauge}
P_{\mu\nu}(k) = -4\pi\eta_{\mu\nu},
\end{eqnarray} with the understanding that the direct Coulomb interaction of charge carriers is excluded from the interaction Hamiltonian \cite{weinberg,landsman}. This so-called Feynman gauge allows simplification of the formula (\ref{expectation3}) in the non-relativistic limit. We recall that the electromagnetic current of charge carriers, $j^{\mu},$ can be obtained by varying the interaction energy with respect to the 4-potential, $\delta W = \int {\rm d}^3\bm{x}j^{\mu}\delta A_{\mu}.$ In the case of non-relativistic charge carriers, the contribution of its spatial components is suppressed by a factor of $v/c$ relative to its temporal component, where $v$ is the charge carrier velocity. Since the photon propagator is diagonal in the Feynman gauge, the temporal and spatial components of $a_{\mu}$ in the pre-exponential factors in Eq.~(\ref{expectation3}) are coupled respectively to the temporal and spatial components of $j^{\mu}$ coming from the interaction vertices. In the non-relativistic limit, therefore, one may set $\hat{\bm{e}} = -\bm{\nabla}\hat{a}_0,$ and neglect $O(v/c)$ terms altogether. Formula (\ref{expectation3}) then simplifies to
\begin{eqnarray}\label{expectation4}
\left\langle\widehat{\Delta U}(t)\widehat{\Delta U}(t')\right\rangle = {\rm tr}\left(\hat{\rho}_0\EuScript{T}_C\hat{\varphi}^{(2)}(t)
\hat{\varphi}^{(1)}(t')\exp\left\{-{\rm i}\int_{C}{\rm d}t\,\hat{w}(t)\right\}\right) - U^2_0,
\end{eqnarray}
where $$\hat{\varphi}(t) = \hat{a}_{0}(t,\bm{x}_1)-\hat{a}_{0}(t,\bm{x}_2).$$

\section{Evaluation of the power spectrum of fundamental $1/f$ noise}\label{evaluation}

\subsection{Feynman diagrams for the leading term of the power spectrum}\label{feynman}

\begin{figure}
\includegraphics[width=1\textwidth]{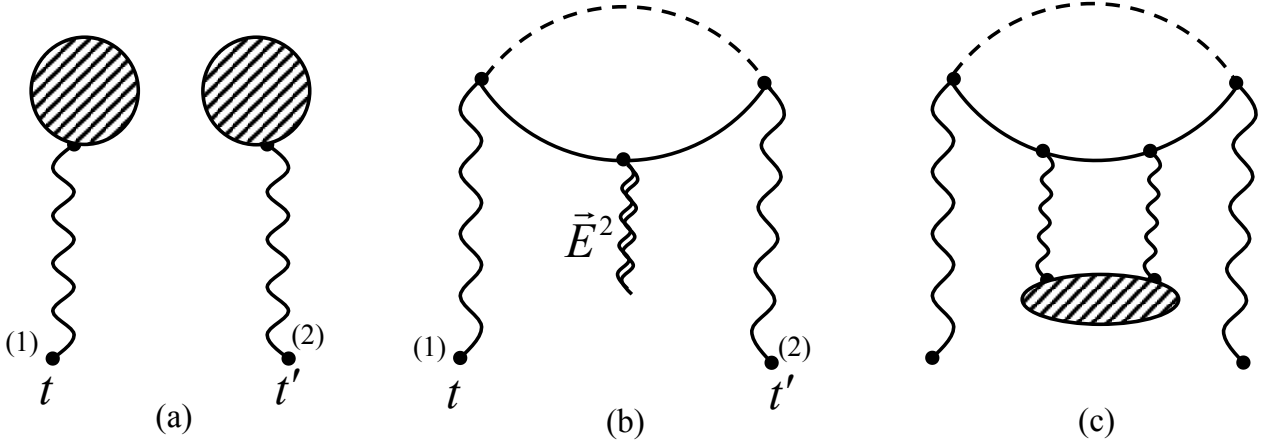}
\caption{Feynman diagrams representing the right hand side of Eq.~(\ref{expectation4}). (a) The disconnected contribution. (b) The lowest-order connected contribution. (c) Diagrams describing particle collisions; the shaded blob is $O(e^2).$ Solid (broken) lines denote part $D_0$ $(\Delta)$ of the charge carrier propagator, ordinary wavy lines -- photon propagators; the double wavy line in (b) designates $\bm{E}^2$.}\label{fig1}
\end{figure}

Among various contributions to the right hand side of Eq.~(\ref{expectation4}) there are terms that factorize into two parts, each of which depends on either $t$ or $t'.$ The corresponding diagrams are disconnected, Fig.~\ref{fig1}(a). Evidently, these terms represent the product $\left\langle\hat{U}(t)\right\rangle\left\langle\hat{U}(t')\right\rangle$, and so their sum just cancels the last term in Eq.~(\ref{expectation4}), provided that all charges present in the system, localised as well as mobile, are taken into account in the shaded blob in Fig.~\ref{fig1}(a). It may be remarked also that irrespective of this cancellation, the disconnected contributions are to be discarded altogether because their sum is symmetric under the interchange $t\leftrightarrow t',$ whereas it was found in Sec.~\ref{masterformula} that the power spectrum of fundamental noise is determined by the antisymmetric part of $\left\langle\widehat{\Delta U}(t)\widehat{\Delta U}(t')\right\rangle$.

\begin{figure}
\includegraphics[width=0.75\textwidth]{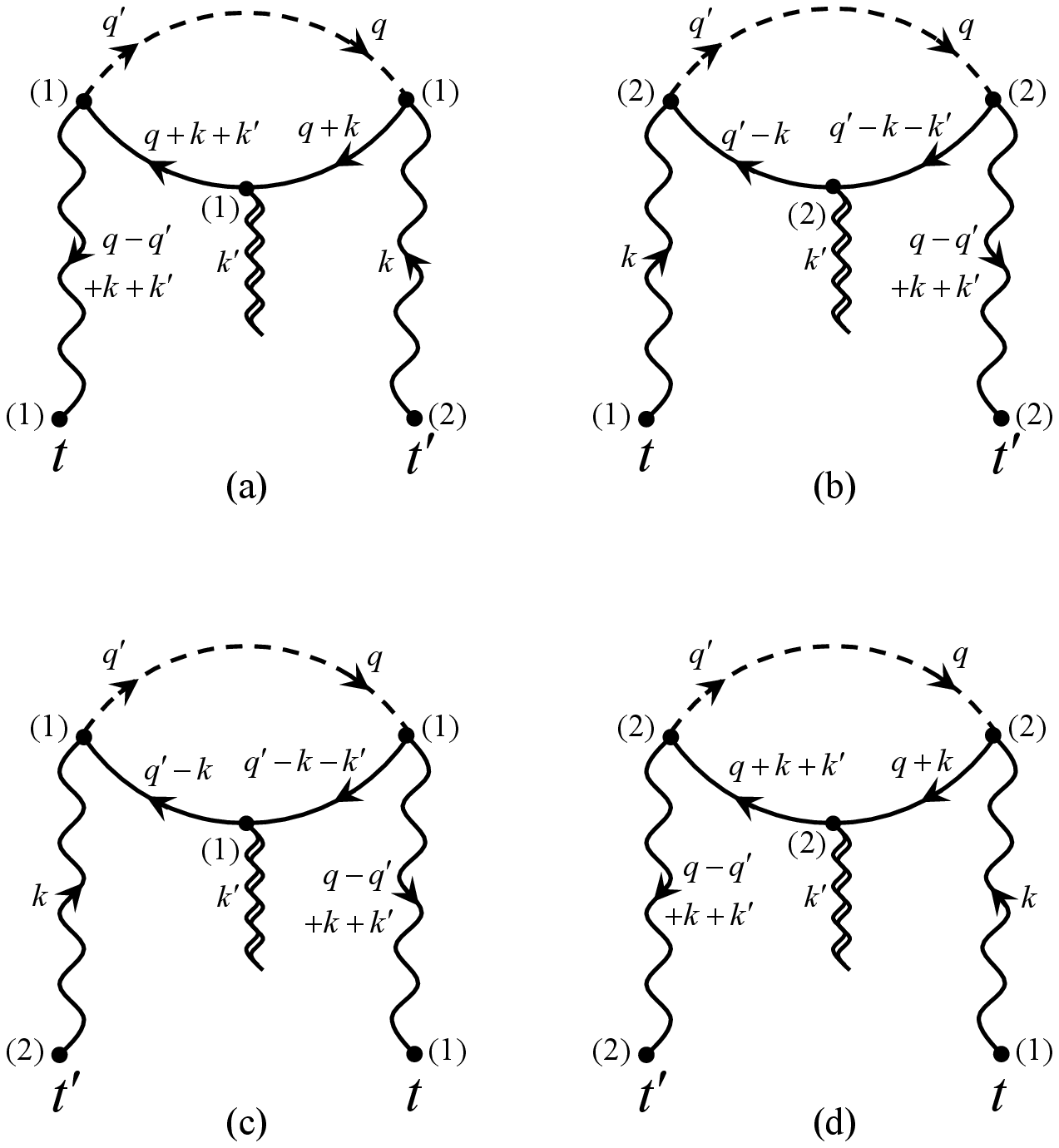}
\caption{Non-vanishing diagrams representing the basic diagram Fig.~\ref{fig1}(b) in the Schwinger-Keldysh formalism. The arrows on the lines show the energy-momentum flow.}\label{fig2}
\end{figure}

In the limit of non-relativistic charge carriers, the interaction Hamiltonian reads
\begin{eqnarray}\label{hamiltonian}
\hat{w}(t) = \int {\rm d}^3\bm{x}\left[e\hat{a}_0(x) + \frac{e^2}{2m}\bm{a}^2(x) \right]\hat{\phi}^{\dagger}(x) \hat{\phi}(x),
\end{eqnarray} where $\bm{a}(x)$ is the vector potential of external electric field,
\begin{eqnarray}\label{avector}
{\bm a}(x) = {\rm i}\bm{E}(\bm{x})\frac{{\rm e}^{{\rm i\lambda t}} - 1}{\lambda}\,, \quad \lambda \to 0.
\end{eqnarray}
This representation of $\bm{a}$ is convenient for the purpose of deriving a Fourier decomposition for the antisymmetric part of $S(t-t')$ with respect to time, which exists according to the discussion in Sec.~\ref{masterformula}. That its spatial Fourier decomposition also exists is plain in view of finiteness of the system spatial extent. Therefore, it is legitimate to go over to the momentum-space representation of Feynman diagrams, which is obtained in the usual way by substituting expressions (\ref{photonprop1}), (\ref{delta}) for the propagators and performing space-time integrations in the interaction vertices. The basic diagram representing the lowest-order nontrivial contribution to $S_F$ is drawn in Fig.~\ref{fig1}(b). First, general comments regarding its structure are in order. As long as the $1/f$ noise is detected in the presence of electric field, any diagram contributing to $S_F$ must involve at least one vertex of the charge-carrier interaction with the external field, so that the lowest-order contribution is $\sim \bm{E}^2$ (the external electric field squared is symbolised in the figure by a double wavy line). In addition to that, there will be two vertices generated by the first term in the integrand of Eq.~(\ref{hamiltonian}), with a photon propagator (wavy line) attached to each, which correspond to pairings of the two $\hat{a}_0$s in the pre-exponent with those from the exponent in Eq.~(\ref{expectation4}). Thus, the lowest-order diagram is of the fourth order with respect to the particle charge $e.$ On the other hand, account of the charge carrier collisions would give rise to terms $O(e^6)$ in $S_F,$ Fig.~\ref{fig2}(c). The easiest way to see this is to formally consider the effect of collisions as an $O(e)$ modification of the external field $\bm{E}$ that acts on the given charge carrier. The fields describing particle interactions are much stronger than $\bm{E},$ rapidly varying in space and time, but the mean value of the net field vanishes. Therefore, it is the field variance that affects $S_F\sim\bm{E}^2,$ so that account of the particle collisions would only add terms $O(e^6)$ indeed. It follows that the lowest-order term of $S_F$ with respect to $\bm{E}$ can be found neglecting particle collisions. But the higher-order contributions cannot be calculated in this way. In fact, the next term of the expansion of $S_F$ in powers of the external electric field is $\sim(\bm{E}^2)^2,$ hence it is also $O(e^6),$ and so keeping this term while neglecting much larger contributions due to the particle collisions would not be legitimate.

Next, the basic diagram necessarily involves a factor of $\Delta(x,y)$ (broken line). Since the other term in the charge carrier propagator, $D_0(x,y),$ as well as the function $G_{\mu\nu}(x,y)$ depend only on $(x-y),$ diagrams without $\Delta$ are spatially homogeneous, and as such are clearly irrelevant to the field produced by a sample of finite volume. The density matrix $\varrho(\bm{q},\bm{q}')$ contained in $\Delta$ ensures that the charge carriers whose fluctuating field is measured are within the sample. By this reason, the external electric field $\bm{E}(x)$ can be considered as $\bm{x}$-independent, that is, as homogeneous in the whole space. The momentum transfer from the field to the charge carriers then vanishes explicitly, as it should once the particle collisions are neglected under the condition of constant current through the sample.

According to the Schwinger-Keldysh rules, each interaction vertex is assigned an index that takes on values 1 and 2. Thus, there are eight different diagrams with the same basic structure of Fig.~\ref{fig1}(b), but only four of them are non-vanishing. This is because a free charge carrier cannot emit a real photon (by virtue of the energy-momentum conservation), and so all interaction vertices in Fig.~\ref{fig1}(b) must be of the same type -- either (1) or (2) (notice also that $D^{(12)}_{0}(q)\equiv 0$). By the same reason, diagrams with more $\Delta$-factors vanish identically. This leaves one with four diagrams shown in Fig.~\ref{fig2}. The choice of independent momentum variables made in this figure proves to be most convenient in the subsequent calculations. Arrows on the lines show the momentum flow; those on the charge carrier propagators are concurrent with the direction from $\hat{\phi}^{\dagger}$ to $\hat{\phi}.$ A line with momentum $k$ coming in (going out) a vertex with 4-coordinates $x$ brings in a factor of ${\rm e}^{-{\rm i}kx}$ (${\rm e}^{{\rm i}kx}$). In particular, $k$ in all cases is assigned to the photon propagators $G^{(12)}$ and $G^{(21)},$ so that it satisfies $k^2=0$ [Cf. Eq.~(\ref{photonprop1})]. According to the conclusion of the preceding paragraph, the 4-momentum associated with the vector potential (\ref{avector}) is $(\lambda,\bm{0}).$ Therefore, the momentum coming in the interaction vertex $\sim \bm{a}^2$ is $k' = (\lambda + \lambda',\bm{0}),$ with each of $\lambda,\lambda'$ being related to one of the factors $\bm{a}$ by the formula (\ref{avector}).

\subsection{Normalization of the one-particle density matrix}\label{normalization}

The function $\varrho(\bm{q},\bm{q}')$ as representing the density matrix of a many-particle system is normalized in the usual way according
$$\iint\frac{{\rm d}^3 \bm{q}}{(2\pi)^3}\frac{{\rm d}^3 \bm{q}'}{(2\pi)^3}\varrho(\bm{q},\bm{q}') = \frac{N}{\Omega}\,,$$
where $N\gg 1$ is the number of charge carriers in the sample of volume $\Omega.$ At the same time, diagrams in Fig.~\ref{fig2} are linear with respect to $\varrho(\bm{q},\bm{q}')$. This does not mean, however, that the power spectrum is proportional to $N,$ because not every contribution to the integral over $\bm{q},\bm{q}'$ in these diagrams also contributes to the power spectrum. As was already mentioned in Sec.~\ref{masterformula}, the terms surviving in the limit $t_m \to \infty$ are those that depend on $t,t'$ only via the difference $(t-t').$ It follows that the time component of $(q-q'+k+k')$ must be equal to that of $k$ [Cf. the 4-momenta assigned to the wavy lines in Fig.~\ref{fig2} and the expression (\ref{photonprop1}) for the photon propagator]. Since $k^{\prime0} = \lambda+\lambda'$ is ultimately set equal to zero, the condition is $q^0=q^{\prime 0}.$ Now, the charge carrier energy in the sample is quantized. To be sure, the spacing of energy levels in a macroscopic sample is negligible in many respects, so that the spectrum can be considered quasi-continuous. But its actual discreteness turns out to be essential in the present context, because the above condition requires that the energies $q^0$ and $q^{\prime 0}$ be the same energy eigenvalue. Therefore, in order to identify the non-vanishing terms in the power spectrum, one has to count the number of contributions satisfying this condition. This task can be greatly simplified by noting that since all thermal as well as particle collision effects have been neglected, the result of this counting must be the same as in the case when the number of filled states equals the number of charge carriers, $N$ (that is, as for an ideal Fermi-gas at zero temperature). We recall that the charge carrier momentum is smeared in each energy eigenstate, and that the set of all momentum eigenstates is complete (as is the set of all energy eigenstates). That is, when the two integration variables $\bm{q},\bm{q}'$ run independently over all momenta, the corresponding energies run over all energy eigenvalues. Hence, for $N$ occupied states, there is a total of $N^2$ pairs of states, of which only $N$ have the same energy. Thus, the fraction of contributions to the power spectrum that satisfy the above condition is $1/N.$ In effect, when using the density matrix $\varrho(\bm{q},\bm{q}')$ for calculating the power spectrum, it is to be understood as normalized by
\begin{eqnarray}\label{normalization1}
\iint\frac{{\rm d}^3 \bm{q}}{(2\pi)^3}\frac{{\rm d}^3 \bm{q}'}{(2\pi)^3}\varrho(\bm{q},\bm{q}') = \frac{1}{\Omega}\,.
\end{eqnarray} It follows that the {\it power spectrum in the approximation considered is independent of the number of charge carriers.} As a notation reminder, the extraction of contributions to $\left\langle\widehat{\Delta U}(t)\widehat{\Delta U}(t')\right\rangle$ that depend only on $(t-t')$ was symbolized in Sec.~\ref{masterformula} by a minus subscript. Thus, this extraction is accomplished by setting $q^0=q^{\prime 0}$ in the Feynman diagrams and normalizing the charge carrier density matrix by Eq.~(\ref{normalization1}).

It is important for the consideration just carried out that the power spectrum definition is based on taking the limit $t_m \to \infty.$ In actual experiments though, one always deals with finite measurement times which are usually taken $t_m \sim 1/\omega,$ and so it is important to assess applicability of the obtained result in such circumstances. Evidently, with $t_m$ of the order of $1/\omega,$ the above energy condition is replaced by $|q^0-q^{\prime 0}| \lesssim \hbar\omega.$ Therefore, the counting of relevant states remains the same until $\hbar\omega$ exceeds the energy spacing. The latter is $1/(D\Omega),$ where $D$ is the density of states of charge carriers in the given material. We see that taking $t_m \sim 1/\omega$ is inconsequential in regard of the noise magnitude at sufficiently low frequencies,
\begin{eqnarray}\label{tmnotinfty}
\omega \lesssim \frac{1}{\hbar D\Omega}\,.
\end{eqnarray}
The density of states varies significantly from one solid to another, but virtually always $D \lesssim 10^{22}$/(eV$\cdot$cm$^3$), while the sample volume in flicker noise studies is normally $10^{-12}$ cm$^3$ to $10^{-8}$ cm$^3.$ Within these limits, therefore, condition (\ref{tmnotinfty}) is well satisfied already for $f\lesssim 1$Hz, but it can be violated at larger frequencies and/or larger samples, in which cases the measured power spectrum would be {\it larger}, roughly by a factor of $(\hbar\omega D\Omega)^2$.

The matrix $\varrho(\bm{q},\bm{q}')$ is customarily expressed via the mixed position-momentum distribution function $f(\bm{r},\bm{Q}),$
\begin{eqnarray}\label{mixed}
\varrho\left(\bm{Q}-\frac{\bm{p}}{2},\bm{Q}+\frac{\bm{p}}{2}\right) = \int_{\Omega} {\rm d}^3\bm{r}{\rm e}^{{\rm i}\bm{p}\cdot\bm{r}}f(\bm{r},\bm{Q}).
\end{eqnarray} With $\varrho(\bm{q},\bm{q}')$ normalized according to Eq.~(\ref{normalization1}), probability distributions for the particle position in the sample or its momentum are obtained by integrating
$f(\bm{r},\bm{Q})$ over all momenta or the sample volume, respectively.

\subsection{Low-frequency asymptotic of the power spectrum}\label{lowfrequency}

On applying the rules formulated in Secs.~\ref{feynman} and \ref{normalization} to diagrams in Fig.~\ref{fig2}, the lowest-order contribution to the function $S(\tau) = \left\langle\widehat{\Delta U}(t'+\tau)\widehat{\Delta U}(t')\right\rangle_- $ takes the form
\begin{eqnarray}\label{lowest}
S(\tau) = &&\frac{{\rm i}(4\pi e^2)^2\bm{E}^2}{2m}\frac{\partial^2}{\partial\lambda\partial\lambda'}\int\frac{{\rm d}^4 k}{(2\pi)^4}\frac{{\rm d}^3 \bm{q}}{(2\pi)^3}\frac{{\rm d}^3 \bm{q}'}{(2\pi)^3}\varrho(\bm{q},\bm{q}')\left[{\rm e}^{{\rm i}\bm{k}\cdot(\bm{x}_1 - \bm{x}_2)} - 1\right] {\rm e}^{{\rm i}(\bm{q}-\bm{q}')\cdot\bm{x}_1 }\nonumber\\&&\times\left\{{\rm e}^{-{\rm i}k^0\tau-{\rm i}k^{\prime0}t}G^{(11)}(q-q'+k+k')D^{(11)}(q+k+k')D^{(11)}(q+k)G^{(12)}(k)
\right.\nonumber\\&& \left. - {\rm e}^{{\rm i}k^0\tau-{\rm i}k^{\prime0}t'}G^{(21)}(k) D^{(22)}(q'-k)D^{(22)}(q'-k-k')G^{(22)}(q-q'+k+k')
\right.\nonumber\\&& \left. + {\rm e}^{-{\rm i}k^0\tau-{\rm i}k^{\prime0}t}G^{(12)}(k)D^{(11)}(q'-k)D^{(11)}(q'-k-k')G^{(11)}(q-q'+k+k')
\right.\nonumber\\&& \left.\left. - {\rm e}^{{\rm i}k^0\tau-{\rm i}k^{\prime0}t'} G^{(22)}(q-q'+k+k')D^{(22)}(q+k+k')D^{(22)}(q+k)G^{(21)}(k)\right\}\right|_{\lambda=\lambda'=0} \nonumber\\ &&+ (\bm{x}_1 \leftrightarrow\bm{x}_2), \nonumber
\end{eqnarray} where ``$+ (\bm{x}_1 \leftrightarrow\bm{x}_2)$'' means that the preceding expression is to be added with $\bm{x}_1$ and $\bm{x}_2$ interchanged. It is perhaps worth reiterating that this Fourier decomposition is only valid for the antisymmetric (that is, odd in $\tau$) part of $S(t-t')$; its symmetric part is kept for a while for notational simplicity, but it will be omitted eventually. It is not difficult to see that the $\lambda$-differentiation of the factors ${\exp}(-{\rm i}k^{\prime0}t),$ ${\exp}(-{\rm i}k^{\prime0}t')$ gives rise to symmetric terms to be omitted. The leading low-frequency term in $S_F(f)$ corresponds to terms in which $\partial^2/\partial\lambda\partial\lambda'$ acts on the charge-carrier propagators $D^{(11)}(q\pm k \pm k')$, $D^{(22)}(q\pm k\pm k'),$ because these approach their poles as $k^0\to 0.$ Since $|\bm{k}| = |k^0|,$ one has $\varepsilon_{\bm{q+k}} = \varepsilon_{\bm{q}} + |k^0|O(|\bm{q}|/m),$ and therefore, in the non-relativistic limit, $D^{(11)}(q+k+k') = {\rm i}/(q^0 + k^0 + \lambda+\lambda' - \varepsilon_{\bm{q+k}})\approx {\rm i}/(k^0+\lambda+\lambda').$ Expanding also ${\rm e}^{{\rm i}\bm{k}\cdot(\bm{x}_1 - \bm{x}_2)}$ to the second order and performing integration over $\bm{k}$ gives
\begin{eqnarray}
&&S(\tau)= \frac{16e^4\bm{E}^2(\bm{x}_1 - \bm{x}_2)^2}{3m}\int_{0}^{\infty}{\rm d}k^0\frac{{\rm e}^{{\rm i}k^0\tau}}{k^0}\int\frac{{\rm d}^3 \bm{q}}{(2\pi)^3}\frac{{\rm d}^3 \bm{q}'}{(2\pi)^3}\varrho(\bm{q},\bm{q}')\frac{{\rm e}^{{\rm i}(\bm{q}-\bm{q}')\cdot\bm{x}_1 } + {\rm e}^{{\rm i}(\bm{q}-\bm{q}')\cdot\bm{x}_2 }}{(\bm{q} - \bm{q}')^2}\,.\nonumber\\
\end{eqnarray} As anticipated, the frequency integral in the antisymmetric part of $S(\tau)$ is convergent, namely, $\int_{0}^{\infty}{\rm d}k^0\sin(k^0\tau)/k^0 = \pi \tau/(2|\tau|).$ The obtained expression for the antisymmetric part of $S(\tau)$ is to be substituted into the right hand side of Eq.~(\ref{sigma}). Using the formulas
\begin{eqnarray}
\int_{-t_m}^{t_m}{\rm d}\tau \frac{\tau}{|\tau|} {\rm e}^{{\rm i}\omega\tau} &=& \frac{2{\rm i}}{\omega}[1- \cos(\omega t_m)], \quad
\frac{1}{t_m}\int_{-t_m}^{t_m}{\rm d}\tau \tau{\rm e}^{{\rm i}\omega\tau} = - \frac{2{\rm i}}{\omega}\cos(\omega t_m) + O(1/t_m), \nonumber
\end{eqnarray} we see that, just like in the above example with $\ln|\tau|$ in the symmetric part of $S(\tau),$ neither of the two integrals in Eq.~(\ref{sigma}) converge, but their difference has a well-defined limit as $t_m\to \infty.$

In practice, the voltage probes are usually aligned parallel to $\bm{E}$ (the noise measured in this configuration is sometimes called longitudinal). In this case $\bm{E}^2(\bm{x}_1 - \bm{x}_2)^2 = U^2_0.$ The remaining momentum integrals are conveniently evaluated by expressing $\varrho(\bm{q},\bm{q}')$ via the function $f(\bm{r},\bm{Q})$ using Eq.~(\ref{mixed}). The integral over $\bm{p}=\bm{q}'-\bm{q}$ in Eq.~(\ref{lowest}) is then a Fourier decomposition of the Coulomb potential. Probability distribution for the particle position in the sample is obtained by integrating $f(\bm{r},\bm{Q})$ over all momenta $\bm{Q} = (\bm{q}+\bm{q}')/2.$ In view of the assumed macroscopic sample homogeneity, $\int{\rm d}^3 \bm{Q}f(\bm{r},\bm{Q})$ is position-independent within the sample (vanishing outside of it). At last, restoring the ordinary units, the power spectrum takes the form
\begin{eqnarray}\label{powerfinal}
S_F(f) = \frac{\varkappa U^2_0}{|f|}\,, \quad \varkappa \equiv \frac{2e^4 g}{\pi m\hbar c^3}\,,
\end{eqnarray} where $g$ is a geometrical factor (it is defined as in Ref.~\cite{kazakov2}) $$g = \frac{1}{3\Omega}\int\limits_{\Omega}{\rm d}^3\bm{r}\left(\frac{1}{|\bm{r}-\bm{x}_1|} + \frac{1}{|\bm{r}-\bm{x}_2|}\right).$$ Thus, the power spectrum of fundamental noise is of the form typical of the observed flicker noise -- it is quadratic with respect to the voltage bias and inversely proportional to frequency.

In some experiments, however, the voltage probes are located differently. In the rather rare case of $(\bm{x}_1 - \bm{x}_2)$ perpendicular to $\bm{E}$, the so-called transverse noise is measured. Specifically, if the current leads and voltage probes are attached to the surface of a rectangular film of length $l$ and width $w,$ then $|\bm{x}_1 - \bm{x}_2| = w,$ $|\bm{E}|l=U_0,$ so that the factor $g$ in the formula (\ref{powerfinal}) is to be replaced by
\begin{eqnarray}\label{powerfinaltr}
g^{\rm tr} = \left(\frac{w}{l}\right)^2\frac{1}{3\Omega}\int\limits_{\Omega}{\rm d}^3\bm{r}\left(\frac{1}{|\bm{r}-\bm{x}_1|} + \frac{1}{|\bm{r}-\bm{x}_2|}\right).
\end{eqnarray}

\subsection{$\gamma \ne 1$: account of the charge carrier--phonon interaction}\label{gammanonunit}

The value of the frequency exponent $\gamma$ is clearly a very important characteristic of the flicker noise spectrum, but it is not always paid due attention in experimental studies. One reason is that its accurate determination requires sufficiently large frequency spans, usually 3--4 decades to achieve an experimental error in $\gamma$ less than $0.05.$ At the same time, a rather strong correlation should be expected between $\gamma$ and the noise magnitude, as suggested by the following simple dimensional reason. As $\gamma$ deviates from unity, parameter $\varkappa$ in Eq.~(\ref{powerfinal}) becomes dimensional, $\varkappa \to \varkappa (f_*)^{\delta},$ where $\delta \equiv \gamma - 1$ and $f_*$ is a constant with the dimension of frequency. Since the deviation must be caused by some physical process in the conductor, $f_*$ is on the order of the corresponding inverse characteristic time, therefore, it is very large when measured in Hertz. If, for instance, the charge carrier mean free time is the relevant time scale ($10^{-14}$ s -- $10^{-13}$ s at room temperature), then the noise magnitude acquires a factor of $10^{13\delta},$ which is $\approx 20$ in the case of $\delta=0.1$  Undoubtedly, this sensitivity of the noise magnitude to $\delta,$ which  can hardly be controlled experimentally, is one of the main reasons why the numerous empirical attempts to come closer to the flicker noise origin have been inconclusive.

One of the mechanisms generating $\delta\ne 0$ is through the charge carrier interaction with acoustic phonons. Consider an acoustic wave propagation in a system comprised of the sample, say a thin film, a substrate it is grown on, and a current source including the wires connecting it to the sample via the current leads. This propagation essentially depends on the acoustic impedance ratios of the system elements. For example, in an idealised case of the current source acoustic impedance exactly equal to that of the sample, acoustic waves go out of the sample without reflections at the current leads. Interaction with the waves affects the charge carrier propagation, and thereby, the $1/f$ noise it produces. In this regard, the long-wavelength phonons turn out to be particularly important. Under the assumption of no reflections, their propagation is as if the system was of infinite extent. In other words, when counting the phonon states and performing summations thereon, $\Omega$ can be considered as large as needed to justify these operations.
   
Of special significance is the long-range piezoelectric interaction exhibited by many semiconductors. Its Hamiltonian reads, in momentum representation, 
\begin{eqnarray}
\hat{w}_{\rm ep} = {\rm i}\sum\limits_{\bm{l}}
\frac{M_{\lambda}(\hat{\bm{l}})}{\sqrt{2\omega_{\bm{l}} \rho_0 \Omega}}\hat{n}(\bm{l})(\hat{a}_{\bm{l}} + \hat{a}^{\dagger}_{-\bm{l}}),
\end{eqnarray} where $\hat{a}_{\bm{l}}$ are the phonon destruction operators, $\omega_{\bm{l}} = u|\bm{l}|,$ $u$ denoting the acoustic wave velocity, $\hat{n}(\bm{l})$ is the Fourier transform of the charge carrier density, $\rho_0$ is the sample density, and $M_{\lambda}(\hat{\bm{l}}) = - M_{\lambda}(-\hat{\bm{l}})$ is the piezoelectric matrix element which only depends on the phonon wavevector direction $\hat{\bm{l}} = \bm{l}/|\bm{l}|$ \cite{mahan}. For the present purposes, it will suffice to have the $(11)$-component of the phonon propagator, which is conventionally defined as $$C^{(11)}(x^0-y^0,\bm{l}) = \left\langle 0 |T(\hat{a}_{\bm{l}} + \hat{a}^{\dagger}_{-\bm{l}})(x^0)(\hat{a}_{-\bm{l}} + \hat{a}^{\dagger}_{\bm{l}})(y^0)|0\right\rangle.$$ Its Fourier transform $$C(l^0,\bm{l}) = \frac{2{\rm i}\omega_{\bm{l}}}{(l^0)^2 - \omega^2_{\bm{l}} + {\rm i}0}\,.$$

\begin{figure}
\includegraphics[width=0.8\textwidth]{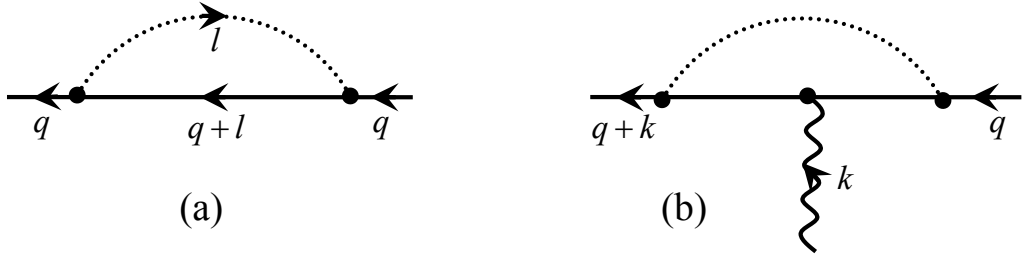}
\caption{Account of the charge carrier interaction with phonons. (a) The lowest-order charge carrier self-energy. (b) The lowest-order vertex correction. Phonon propagators are depicted by dotted lines.}\label{fig3}
\end{figure}

The interaction with phonons modifies the charge carrier propagator
$$D^{(11)}_{0}(q) = \frac{{\rm i}}{q^0 - \varepsilon_{\bm{q}} + {\rm i}0} \to \EuScript{D}^{(11)}_{0}(q) = \frac{{\rm i}}{q^0 - \varepsilon_{\bm{q}} - \sigma(q) + {\rm i}0}\,,$$ where the charge carrier self-energy $\sigma(q)$ in the second order of perturbation theory is represented by the diagram in Fig.~\ref{fig3}(a). Since the charge carrier propagators in the diagrams of Fig.~\ref{fig2} have the 4-momenta off the mass shell, so do the external lines in Fig.~\ref{fig3}(a), and therefore, the vertices in this diagram must be of the same type -- both are either $(1)$ or $(2),$ according to the type of the propagator which the self-energy correction refers to. But since the energy-momentum exchange with phonons can move the charge carrier on the mass shell, the pole structure of the inner charge carrier propagator in Fig.~\ref{fig3}(a) becomes important, that is, the state filling needs to be taken into account. Assuming for definiteness an empty band, one has
\begin{eqnarray}\label{selfenergy}
\sigma(q) = \frac{{\rm i}}{\rho_0}\int\frac{{\rm d}^3\bm{l}}{(2\pi)^3}\int_{-\infty}^{+\infty}\frac{{\rm d}l^0}{2\pi}\frac{|M_{\lambda}(\hat{\bm{l}})|^2}{[(l^0)^2 - \omega^2_{\bm{l}} + {\rm i}0][q^0 + l^0 - \varepsilon_{\bm{q+l}}+ {\rm i}0]}\,.
\end{eqnarray} The $l^0$-integration amounts to taking the residue of the integrand at $l^0 = - \omega_{\bm{l}}.$ To be more specific, consider the simplest case of parabolic dispersion of the charge carrier energy, $\varepsilon_{\bm{q}} = \varepsilon_0 + \bm{q}^2/2m,$ where $\varepsilon_0$ is an arbitrary constant. As we only take into account contributions to $S_F$ which are independent of the charge carrier velocity, $\bm{q}$ is to be set equal to zero, so that $\varepsilon_{\bm{q+l}} = \varepsilon_0 + \bm{l}^2/2m,$ and integration over $\bm{l}$ yields
\begin{eqnarray}\label{sigma1}
\sigma(q^0,\bm{0}) = \frac{M^2_{\lambda}}{(2\pi)^2\rho_0u^3}(q^0 - \varepsilon_0)\ln\frac{|q^0 - \varepsilon_0|}{l_m u} + \Delta\varepsilon_0,
\end{eqnarray}
where $M^2_{\lambda}$ is the angular mean of $|M_{\lambda}(\hat{\bm{l}})|^2,$ $\Delta\varepsilon_0$ is a contribution  independent of the charge carrier energy, and $l_m$ is a maximal phonon momentum. As the preceding analysis has shown, the $1/f$ asymptotic of the power spectrum is determined by the properties of the charge carrier propagator near its pole, wherein $|q^0 - \varepsilon_0| = |\omega| \lll l_m u.$ Therefore, within the logarithmic accuracy, $l_m$ can be taken direction-independent and equal to $2\pi/d,$ $d$ denoting the lattice constant. Also, the constant $\Delta\varepsilon_0$ can be omitted, as on substituting Eq.~(\ref{sigma1}) into $\EuScript{D}$ it amounts to a redefinition of $\varepsilon_0.$

The interaction with phonons affects electromagnetic interactions of the charge carriers as well. The lowest-order vertex correction is drawn in Fig.~\ref{fig3}(b), and its evaluation is quite similar to that just performed for the self-energy. The result is that when a charge carrier with momentum $q$ on the mass shell exchanges momentum $k$ with the electromagnetic field, the vertices (\ref{hamiltonian}) are multiplied by a factor $[1+\Gamma(q,k)],$ where
\begin{eqnarray}\label{gamma}
\Gamma(q,k) = - \frac{M^2_{\lambda}}{(2\pi)^2\rho_0u^3}\ln\frac{|k^0|}{l_m u}\,.
\end{eqnarray} It is not difficult to check that in the diagrams of Fig.~\ref{fig2}, this correction to any of the two side vertices just cancels the contribution (\ref{sigma1}) to the charge carrier propagator connecting it to the central vertex. The phonon correction to the central vertex itself is slightly different from Eq.~(\ref{gamma}), for the charge carrier lines attached to it are both off the mass shell,
\begin{eqnarray}\label{gamma1}
\Gamma(q+k,k') &=&
- \frac{M^2_{\lambda}}{(2\pi)^2\rho_0u^3}\left(\ln\frac{|k^0 + k^{\prime0}|}{l_m u} + \frac{k^0}{k^{\prime0}}\ln\left|1+ \frac{k^{\prime0}}{k^0}\right| \right)\nonumber\\
&=& - \frac{M^2_{\lambda}}{(2\pi)^2\rho_0u^3}\left(\ln\frac{|k^0 + k^{\prime0}|}{l_m u} + 1 - \frac{1}{2}\frac{k^{\prime0}}{k^0} + \frac{1}{3}\left(\frac{k^{\prime0}}{k^0}\right)^2 + \dots\right).
\end{eqnarray} However, the additional terms do not change the general structure of Eq.~(\ref{powerfinal}), giving rise only to relatively small corrections to $\varkappa.$ On the other hand, the logarithmic term is important as it changes the analytical structure of diagrams.
Denoting
\begin{eqnarray}\label{delta1}
\delta = \frac{M^2_{\lambda}}{(2\pi)^2\rho_0u^3},
\end{eqnarray}
the product of the factor $[1+\Gamma(q+k,k')]$ with the propagator $D(q+k+k')$ can be rewritten, within the second-order accuracy, as
$$[1+\Gamma(q+k,k')]D^{(11)}_{0}(q+k+k') = {\rm i}\frac{1 - \delta \ln\frac{|k^0 + k^{\prime0}|}{l_m u}}{k^0 + k^{\prime0} + {\rm i}0} =  \frac{{\rm i}}{\left(k^0 + k^{\prime0} + {\rm i}0\right)}\left|\frac{l_m u}{k^0 + k^{\prime0}}\right|^{\delta}.$$ This form makes it evident that the effect of the logarithmic term is to replace Eq.~(\ref{powerfinal}) with
\begin{eqnarray}\label{powerphonon}
S_F(f) = \frac{\varkappa U^2_0}{|f|^{\gamma}}\,, \quad \varkappa \equiv \frac{2e^4 (f_*)^{\delta}g}{\pi m\hbar c^3}\,, \quad \gamma = 1 + \delta, \quad f_* = u/d.
\end{eqnarray}
On restoring the ordinary units, a factor of $\hbar$ is to be inserted in $\delta,$
\begin{eqnarray}\label{delta2}
\delta = \frac{M^2_{\lambda}}{(2\pi)^2\hbar\rho_0u^3},
\end{eqnarray}
Thus, the non-analyticity of $\Gamma$ results in a shift of the frequency exponent from unity. That the shift is positive is directly related to the fact that it is vacuum parts of the charge carrier propagators that appear in Eq.~(\ref{selfenergy}). It is not difficult to check that the sign of $\sigma(q)$ would be opposite if the level with the given $\bm{q}$ was completely filled. In general, therefore, the right hand side of Eq.~(\ref{powerphonon}) is smeared with respect to $\delta.$ It is to be recalled also that the expression (\ref{delta1}) is valid under the assumption of perfect acoustic impedance match between the sample and its surroundings. In the opposite case of complete acoustic reflection, the long-wavelength phonon modes are cut off, so that $\gamma=1.$

\section{Comparison with experiment}\label{comparison}

Even without a detailed analysis, one can assert that in the formal limit of ever-increasing measurement accuracy, the voltage variance grows without bound, because so does the system state variation. Evidently, no valuable information about the system can be obtained this way. In the opposite limit of vanishing measuring device effect, which is unattainable either, the system state remains fixed during the measurement, but the accuracy of any of its outcomes is determined entirely by the uncertainty relation proper to the given state. Despite its unattainability, it is the latter case which is relevant to practice. In fact, minimization of the influence of the measuring apparatus on the system under study is a common strategy of any experiment. Therefore, the noise levels observed in sufficiently clean samples, such that conventional noise sources are all eliminated, are naturally expected to be not very far from the lower bound set by the quantum indeterminacy. To see how far they actually are from the values (\ref{powerfinal}), (\ref{powerfinaltr}), (\ref{powerphonon}), we consider below several typical cases which are also quite revealing on their own in regard of the problems they pose to the traditional view on the flicker noise as originating from the conductance fluctuations.

\subsection{Noise in InGaAs quantum wells}

\subsubsection{Longitudinal noise}

As a first example, we take Ref.~\cite{chenaud2016} reporting noise measurements in In$_{1-x}$Ga$_x$As quantum wells of significantly different sizes. The charge carriers in this case are electrons ($n$) and holes ($p$). Since $\varkappa\sim 1/m,$ it is sufficient to consider the lightest charge carriers whose masses are $m_n = 0.06m_0$, $m_p = 0.09m_0,$ respectively, where $m_0$ is the free electron mass (these are approximate values, as the masses depend on the sample thickness, composition {\it etc.}). With one exception, the spectra measured in Ref.~\cite{chenaud2016} had the frequency exponents very close to unity, and so it is natural to begin with Eq.~(\ref{powerfinal}). The sample dimensions and experimental results taken from figures~4--7 of Ref.~\cite{chenaud2016} are compared in Table~\ref{table1} with the values computed according to Eq.~(\ref{powerfinal}).
\begin{table}
\begin{tabular}{ccccc||c||c}
\hline\hline
{\rm sample}
  & \hspace{0,1cm} $w,$ $\mu$m \hspace{0,1cm}
  & \hspace{0,1cm} $l,$ $\mu$m \hspace{0,1cm}
  & \hspace{0,1cm} $a,$ nm \hspace{0,1cm}
  & $g,$ cm$^{-1}$
  & \hspace{0,5cm} $\varkappa_{\rm th}$ \hspace{0,5cm}
  &  \hspace{0,5cm} $\varkappa_{\rm exp}$  \hspace{0,5cm} \\
\hline
V1& 1 & 2.2 & 10 & 9630 & $3.5\times 10^{-10}$ & $1.75\times 10^{-9}$\\
V1.5& 1.5 & 3.3 & 10 & 6420 & $2.3\times 10^{-10}$ & $4.5\times 10^{-10}$\\
V2& 2 & 4 & 10 & 5140 & $1.9\times 10^{-10}$ & $3.1\times 10^{-10}$\\
V5& 5 & 20 & 20 & 1260 & $4.6\times 10^{-11}$ & $1.5\times 10^{-9}$ \\
V80& 80 & 300 & 20 &  80  & $1.9\times 10^{-12}$ & $4.1\times 10^{-12}$\\
\hline\hline
\end{tabular}
\caption{Comparison of the measured ($\varkappa_{\rm exp}$) and calculated ($\varkappa_{\rm th}$) noise magnitude in various InGaAs samples. Also given are the sample width ($w$), length ($l$) and thickness ($a$).} \label{table1}
\end{table}
It is seen that except for the sample V5, the measured noise levels are only a few times as high as the lower bound set by quantum indeterminacy. As was discussed in Sec.~\ref{gammanonunit}, the calculational accuracy is rather sensitive to the error in $\delta,$ which in the present case is around $0.02\div0.03$ \cite{chaubet}. Taking into account that in the given material, $d\approx 5\times 10^{-8}$cm, $u =(2.5\div5)\times 10^5$cm/s, this gives a factor of $(f_*)^{0.2}\approx 2$ in the noise magnitude. This leads one to conclude that within this accuracy, the observed noise is nearly minimal. To be sure, there can be various reasons for the noise level to exceed the minimum, but it appears that the exceptionally large value of $\varkappa_{\rm exp}$ detected in sample V5 can be to a large extent attributed to the deviation of the measured $\delta$ from unity, which is noticeable just in this case. A least square fit of the measured spectrum yields $\delta \approx 0.05$ \cite{chaubet}, which magnifies the theoretical value of $\varkappa$ by a factor of four.

Although Eq.~(\ref{delta2}) was obtained neglecting thermal effects, whereas the experiments \cite{chenaud2016} were carried at room temperature, it is of interest to estimate the theoretical maximum of $\delta.$ $M^2_{\lambda}$ is customarily evaluated as $(eh_{14})^2,$ where the piezoelectric constant $h_{14} = -1.4\times 10^9$ V/m. The lowest acoustic wave velocity, $u = 2.5\times 10^5$ cm/s, is for propagation in the direction [110]; substitution of these figures together with $\rho_0 = 5.3$ g/cm$^3$ in Eq.~(\ref{delta2}) gives $\delta = 0.14.$ The figures used are for samples with low In fraction; in the case $x=0.47,$ which is of special practical interest, $\delta$ drops to $0.09$.

\subsubsection{Transverse noise}

Ref.~\cite{chenaud2016} is one of a handful of papers that deal with the transverse noise, that is, voltage fluctuations in the direction perpendicular to the current flow in the sample \cite{hawkins1970,hawkins1971,kleinpenning}. Within the conventional model of flicker noise as originating from conductance fluctuations, the theoretical prediction \cite{kleinpenning} (based on a phenomenological inclusion of the $1/f$ term into the fluctuation spectrum) is that for a given electric filed, the noise power density is independent of the distance $|\bm{x}_1 - \bm{x}_2|$ between the voltage probes. On the other hand, according to the present theory, $S_F(f)$ is proportional to this distance squared [Cf. Eq.~(\ref{lowest})]. The computed and measured values of $\varkappa$ are summarized in Table~\ref{table2}. First of all, comparing the last columns in Tables \ref{table1} and \ref{table2}, one readily concludes that the observed levels of transverse noise in all samples are significantly lower than the corresponding levels of the longitudinal noise. Furthermore, it is seen that the reduction by a factor of $(w/l)^2$ predicted by Eq.~(\ref{powerfinaltr}), which varies in the range 1/16 to 1/4, brings $\varkappa_{\rm th}$ to values which are roughly in the same ratios to the measured values as in the case of longitudinal noise.

\begin{table}
\begin{tabular}{ccccc||c||c}
\hline\hline
{\rm sample}
  & \hspace{0,1cm} $w,$ $\mu$m \hspace{0,1cm}
  & \hspace{0,1cm} $l,$ $\mu$m \hspace{0,1cm}
  & \hspace{0,1cm} $a,$ nm \hspace{0,1cm}
  & $g^{\rm tr},$ cm$^{-1}$
  & \hspace{0,5cm} $\varkappa_{\rm th}$ \hspace{0,5cm}
  &  \hspace{0,5cm} $\varkappa_{\rm exp}$  \hspace{0,5cm} \\
\hline
V1& 1 & 2.2 & 10 & 1990 & $7.2\times 10^{-11}$ & $2.4\times 10^{-10}$\\
V1.5& 1.5 & 3.3 & 10 & 1330 & $4.8\times 10^{-11}$ & $1.3\times 10^{-10}$\\
V2& 2 & 4 & 10 & 1280 & $4.8\times 10^{-11}$ & $5.9\times 10^{-11}$\\
V5& 5 & 20 & 20 & 80 & $2.9\times 10^{-12}$ & $7.0\times 10^{-11}$ \\
V80& 80 & 300 & 20 &  6  & $1.4\times 10^{-13}$ & $3.7\times 10^{-13}$\\
\hline\hline
\end{tabular}
\caption{Same as in Table~\ref{table1}, but for transverse noise.} \label{table2}
\end{table}

\subsection{Huge noise in high-$T_c$ superconductors}\label{hightc}

It is of considerable interest to make a similar comparison for high-$T_c$ superconductors, as they are known to exhibit anomalously high levels of $1/f$ noise. It so happened that the  noise in these materials was first measured in samples of sizes unusually large for flicker noise studies -- the linear sample dimensions were several millimeters \cite{testa}. Subsequent measurements in much smaller samples gave much lower noise levels though. This issue was considered in detail in Ref.~\cite{kazakov3} where it was demonstrated that the anomaly in the noise level is spurious; it is caused by an inappropriate normalization of the power spectra using Hooge's formula \cite{hooge1} according to which $S(f)\sim 1/\Omega,$ whereas actually $S(f)$ is inversely proportional to the {\it linear} sample size. Here we will only show on a couple of examples that the ``huge'' noise is in fact not far from the minimum given by Eq.~(\ref{powerphonon}). 

The work \cite{song2} reports $1/f$-noise measurements in bulk samples of YBa$_2$Cu$_3$O$_y.$ The samples used were single crystals with $l = w = 0.2$ cm, $a = 0.01$ cm. The authors give the value $1.06 \pm 0.1$ for the frequency exponent. The charge carriers in this case are holes. A precise determination of the hole effective mass in this material is difficult; the experiments \cite{padilla,minami} suggest that at low temperatures, $m_p \approx (2\div 3)m_0,$ and that it is nearly constant in the range of $y$s for which the material exhibits superconductivity, that is, for sufficiently small oxygen deficiency. In the work under consideration, such is sample B, and the value of $\varkappa$ measured near the superconducting transition is, as read off from Fig.~2 of Ref.~\cite{song2},  $\varkappa_{\rm exp} = 3\times10^{-14}$ Hz$^{0.06}$. On the other hand, Eq.~(\ref{powerphonon}) with $m_p = 3m_0$ yields $\varkappa_{\rm th} = 2.2\times10^{-14}$ Hz$^{0.06}$. It is worth noting that the value of the Hooge parameter found by the authors for this sample is $2.1\times 10^3,$ that is, six orders of magnitude higher than the canonical value $2\times 10^{-3}$ \cite{hooge1}.

In 1994, a systematic investigation of flicker noise in thin films of YBa$_2$Cu$_3$O$_y$ was undertaken in order to determine its dependence on the oxygen content, $y$ \cite{liu}. The Hooge parameter was found to be $\approx 14.$ This is still large compared to pure metals, but several orders of magnitude smaller than those reported previously for the compound, and the authors attributed this reduction to the quality of their thin films. All samples had $a = 8.5\times10^{-6}$ cm, $l = 0.5$ cm, and $w = 0.07$ cm. The noise power was found to exhibit a sharp minimum at $y\approx 6.5,$ in which case the measurements gave $\varkappa_{\rm exp} = 3\times 10^{-15}.$ On the other hand, substitution of the sample dimensions together with $m_p = 3m_0$ as before in Eq.~(\ref{powerfinal}) gives $g=6$~cm$^{-1}$ and the theoretical minimum $\varkappa_{\rm th} = 2.3\times 10^{-15}.$

\section{Discussion and conclusions}\label{discussion}

The uncertainty principle is a fundamental limitation imposed by quantum theory on any measurement process. We have shown that the quantum uncertainty in the electric voltages measured across a conducting sample at different instants sets a lower bound on the power spectrum of voltage fluctuations. Its low-frequency asymptotic turned out to be typical of the observed flicker noise spectra -- it is quadratic with respect to the voltage bias and inversely proportional to frequency. Thus, one can say that flicker noise is as fundamental as the uncertainty principle.

Quantum indeterminacy dictated by the uncertainty principle forced us to abandon the familiar relationship between the voltage autocovariance and power spectrum, the Wiener-Khinchin theorem. The assumption that the measurement process yields a more or less accurate voltmeter reading at any instant implies an accompanying system state variation coordinated with the measurement accuracy. These variations driven by the measuring device make the process non-stationary. On the other hand, if one wants to keep the system state fixed, the voltage fluctuation cannot be defined for all times in principle. The basic assumptions of the Wiener-Khinchin theorem are thus violated in either case.

An essential difference between the voltage autocovariance and the power spectrum is that, although both are dependent on the measuring device operation, the latter is free of the operator ordering ambiguity. It is this fact that allowed us to find its lower bound without the need to explicitly include measuring device into consideration in order to follow the system state evolution. By this reason, the voltage power spectrum as defined by Eq.~(\ref{power}) ({\it not} as the Fourier transform of autocovariance) turns out to be a notion more fundamental than the autocovariance.

As shown in Sec.~\ref{comparison} by comparing Eqs.~(\ref{powerfinal}), (\ref{powerfinaltr}), (\ref{powerphonon}) with the experimental data, the observed flicker noise levels are in all cases only a few times higher than the theoretical minimum. At the same time, as discussed in Sec.~\ref{gammanonunit}, because of the sensitivity of the noise magnitude to the value of the frequency exponent, the calculated noise levels are actually order-of-magnitude estimates. In these circumstances, a conclusion that can be drawn from the comparison is that within this accuracy, the observed noise is nearly minimal. To be sure, this does not exclude the possibility that the noise level might be lower, but is raised to the observed level by an independent measurable effect. In fact, the photon heat bath contribution to the $1/f$ noise in high-temperature superconductors at room temperature is of the same order as the vacuum contribution \cite{kazakov3}.

The existing order-of-magnitude accuracy, which might be considered as disastrous in other areas of science, is yet sufficient to make further important conclusions. First of all, the significant drop of the noise levels on switching from the longitudinal to the transverse configuration, observed in Ref.~\cite{chenaud2016}, takes place without exception in all samples studied. This rules out the conventional interpretation of $1/f$ noise as a result of fluctuations in the sample conductance, which predicts the same noise level in the two configurations. On the other hand, the amount of drop is well correlated with the law $S_F(f)\sim (w/l)^2$ predicted by Eq.~(\ref{powerfinaltr}). From a more general standpoint, the dependence of $S(f)$ on the sample dimensions as given by this equation naturally resolves the puzzle of inexplicably high values of noise observed in comparatively large samples. Specifically, in the case of high-$T_c$ superconductors, the noise magnitude in millimeter-size samples, normalized according to the Hooge's formula, was found to be $10^6$ to $10^{10}$ as high as in micrometer-size thin films. Why samples of comparable quality exhibit such a huge difference in noise levels can be explained only if one admits that the scaling $S(f)\sim 1/\Omega$ implied by Hooge's formula is incorrect. In fact, comparison made in Sec.~\ref{hightc} demonstrates that when analyzed on the basis of Eq.~(\ref{powerphonon}), the noise observed in millimeter-size high-$T_c$ superconductors is as ordinary as in the other instances considered.

\acknowledgments{I am indebted to Christophe Chaubet and his colleagues at the University of Montpellier (France) for providing experimental details and discussion of Ref.~\cite{chenaud2016}.}

\pagebreak

\end{document}